# Electromagnetic energy storage and power dissipation in nanostructures

J. M. Zhao[1,2] and Z. M. Zhang[1*]

[1] *George W. Woodruff School of Mechanical Engineering*
*Georgia Institute of Technology, Atlanta, Georgia, USA*

[2] *School of Energy Science and Engineering*
*Harbin Institute of Technology, Harbin, People's Republic of China*

**Abstract**

The processes of storage and dissipation of electromagnetic energy in nanostructures depend on both the material properties and the geometry. In this paper, the distributions of local energy density and power dissipation in nanogratings are investigated using the rigorous coupled-wave analysis. It is demonstrated that the enhancement of absorption is accompanied by the enhancement of energy storage both for material at the resonance of its dielectric function described by the classical Lorentz oscillator and for nanostructures at the resonance induced by its geometric arrangement. The appearance of strong local electric field in nanogratings at the geometry-induced resonance is directly related to the maximum electric energy storage. Analysis of the local energy storage and dissipation can also help gain a better understanding of the global energy storage and dissipation in nanostructures for photovoltaic and heat transfer applications.

______________________________________________________________
*Corresponding author.
*E-mail addresses: jmzhao@hit.edu.cn (JMZ), zhuomin.zhang@me.gatech.edu (ZMZ)*

**1 Introduction**

Knowledge of the local electromagnetic energy storage and power dissipation is very important to the understanding of light-matter interactions and hence may facilitate structure optimization for applications in energy harvesting, optical heating, photodetection and radiative properties tuning based on nanostructures in the fields of nanophotonics [1], photovoltaics [2], metamaterials [3], and near-field thermal radiation [4]. The electromagnetic energy storage and power dissipation in nanostructures rely both on the materials properties and on the structure geometry. The effect of materials optical property on energy storage and power dissipation density has been studied by many researchers, including early works by Loudon [5], Barash and Ginzburg [6], Brillouin [7], and Landau [8], and more recent works by Ruppin [9], Shin et al. [10], and Vorobyev [11, 12] to name a few. Expressions of the time-averaged energy density [5, 9, 11], instantaneous energy density [10], and power dissipation density for lossy dispersive media have been developed. It is noted that the energy density formulae rely on microphysical parameters of the specific material's dielectric function model, such as Drude model or Lorentz model; although the time-averaged power dissipation density depends only on the imaginary part of the dielectric function and the electric field for a nonmagnetic material.

Though the electromagnetic energy density for bulk materials have been studied by many researchers, the geometric effect in the local energy storage distribution in nanostructures has not been extensively investigated. Meanwhile, local power dissipation analysis has recently been demonstrated to be very useful in understanding the heat generation in nanoparticles [13, 14] and metamaterials [15, 16], and in analyzing the efficiency of photovoltaic solar cells [17-20]. It is anticipated that power dissipation and energy storage are closely related, since both the electric and magnetic fields can store energy but, for nonmagnetic materials, only electric current can result in power dissipation, understanding such relationship may be desirable and offer deep insight of the radiative properties of nanostructures. Hence, a combined analysis of local energy density and power dissipation distributions in nanostructures may help explore their relation as well as the influence of geometric structures and resonance conditions. Because energy is additive, global energy storage and power dissipation can be obtained by integrating the local energy storage and power dissipation, respectively. The objective of the present work is to develop the formulae and procedure for investigating the locally enhanced energy storage and absorption in nanostructures.

In this paper, the formulae of the time-averaged energy density and power dissipation are revisited and summarized. The rigorous-coupled wave analysis (RCWA) method is applied to accurately solve for the electric and magnetic field distributions in nanostructures. Combined with the formulae based on microphysical parameters and dielectric function models, the distributions of local energy storage and local power dissipation in the nanogratings are obtained. A combined analysis of local energy storage and power dissipation in a simple metallic grating support magnetic polariton and a grating-enhanced solar cell structure are studied as examples.

**3 Energy density formula for lossy dispersive media revisited**

**3.1 Power dissipation in lossy dispersive media**

The Poynting theorem is a relation derived from Maxwell's equations that describes the energy balance, which can be written as [21]

$$-\nabla \cdot \mathbf{S} = \mathbf{J} \cdot \mathbf{E} + \mathbf{J}_h \cdot \mathbf{H} \tag{1}$$

where $\mathbf{S} = \mathbf{E} \times \mathbf{H}$ is the Poynting vector, $\mathbf{E}$ and $\mathbf{H}$ are the electric and magnetic fields, respectively, $\mathbf{J} = \mathbf{J}_f + \frac{\partial \mathbf{D}}{\partial t}$ is the combination of the conduction current $\mathbf{J}_f$ and the displacement current that is the time derivative of the electric displacement $\mathbf{D}$, $\mathbf{J}_h = \frac{\partial \mathbf{B}}{\partial t}$ is the magnetic current, and $\mathbf{B}$ is the magnetic induction. The term $-\nabla \cdot \mathbf{S}$ describes the amount of electromagnetic power flows into a differential volume. The inflow energy is partly stored and partly dissipated. As such, Eq. (1) can be rewritten as

$$-\nabla \cdot \mathbf{S} = \frac{\partial}{\partial t}(u_e + u_h) + (w_e + w_h) \tag{2}$$

where $u_e$ and $u_h$ denote the electric and magnetic energy storage density, respectively, and $w_e$ and $w_h$ denote the electric and magnetic power dissipation per unit volume, respectively.

For a monochromatic wave, the real fields and the complex fields amplitude are related by $\mathbf{E} = \mathrm{Re}\left[\tilde{\mathbf{E}}\exp(j\omega t)\right]$ and $\mathbf{H} = \mathrm{Re}\left[\tilde{\mathbf{H}}\exp(j\omega t)\right]$. Throughout this paper, the time dependency of $\exp(j\omega t)$ is used in which $j = \sqrt{-1}$ and $\omega$ is the angular frequency and $t$ is the time. Even for monochromatic wave excitation, the identification of energy storage term and power dissipation term from $\mathbf{J} \cdot \mathbf{E}$ and $\mathbf{J}_h \cdot \mathbf{H}$ will rely on specific material model [5, 9, 10], such as the Drude model or the Lorentz model. The only exception is the time-averaged power dissipation density, which can be derived by time averaging Eq. (1) and Eq. (2) such that

$$\langle w_e \rangle = \langle \mathbf{J} \cdot \mathbf{E} \rangle = \frac{1}{2}\omega\varepsilon_0\varepsilon''|\tilde{\mathbf{E}}|^2 \quad \text{and} \quad \langle w_h \rangle = \langle \mathbf{J}_h \cdot \mathbf{E} \rangle = \frac{1}{2}\omega\mu_0\mu''|\tilde{\mathbf{H}}|^2 \tag{3}$$

Here, $\langle \ \rangle$ denotes the time-average operator, $\tilde{\mathbf{E}}$ and $\tilde{\mathbf{H}}$ are the complex amplitude of the electric and magnetic fields, $\varepsilon_0$ and $\mu_0$ are the vacuum permittivity and permeability, $\varepsilon''$ and $\mu''$ are the imaginary part of the relative permittivity ($\tilde{\varepsilon} = \varepsilon' - j\varepsilon''$) and relative permeability ($\tilde{\mu} = \mu' - j\mu''$), respectively.

### 3.2 Energy storage in materials described by a Lorentz oscillator model

For ideal dielectrics, the formulae for calculating time-averaged electric and magnetic energy storage densities are well known [21], that is,

$$\langle u_e \rangle = \frac{1}{4} \varepsilon_0 \varepsilon \left| \tilde{\mathbf{E}} \right|^2 \quad \text{and} \quad \langle u_h \rangle = \frac{1}{4} \mu_0 \mu \left| \tilde{\mathbf{H}} \right|^2 \tag{4}$$

The calculation of energy storage density in lossy dispersive media is more complicated. Until recently, the most well-developed theory for energy storage density calculation is for the Lorentz model, which is one of the most basic material models and can be applied to model a large group of real materials [22]. It is noted that the Drude model can be viewed as a special case of the Lorentz model with a zero resonance frequency. There are also some energy density formulae derived based on special material model for metamaterials [23-26], which is out of the scope of the present paper.

For a single Lorentz oscillator, the dynamic equation of a bound charge can be written as

$$m \frac{d^2 \mathbf{r}}{dt^2} + m\Gamma \frac{d\mathbf{r}}{dt} + k\mathbf{r} = q\mathbf{E} \tag{5}$$

where $\mathbf{r}$, $m$, $\Gamma$, $k$, and $q$ denote the position, mass, damping coefficient, spring stiffness, and the charge, respectively. The energy stored by the Lorentz oscillator is the sum of the kinetic energy and the potential energy. Alternatively, Eq. (5) can be written into an equivalent RLC circuit form in the frequency domain as

$$\left( j\omega \rho_L + \rho_R + \frac{1}{j\omega \rho_C} \right) \tilde{\mathbf{J}}_b = \tilde{\mathbf{E}} \tag{6}$$

where $\rho_L = 1/(\varepsilon_0 \omega_p^2)$ is an equivalent inductance, $\rho_R = \Gamma/(\varepsilon_0 \omega_p^2)$ is an equivalent resistance, $\rho_C = \varepsilon_0 \omega_p^2 / \omega_0^2$ is an equivalent capacitance, $\omega_p = q\sqrt{N/(\varepsilon_0 m)}$ is the plasma frequency, and $\omega_0 = \sqrt{k/m}$ is the resonance frequency, and $\tilde{\mathbf{J}}_b = j\omega Nq\tilde{\mathbf{r}}$ is the current density for bound charges, in which $N$ is the number density of oscillators. Note that the equivalent inductance, resistance, and capacitance are defined per unit length and per unit cross sectional area of the current. The relative permittivity $\tilde{\varepsilon}_b$ for the bound charges can therefore be obtained as

$$\tilde{\varepsilon}_b = \frac{\omega_p^2}{\omega_0^2 - \omega^2 + j\omega\Gamma} \tag{7}$$

In the equivalent circuit form, the energy storage by kinetic energy and potential energy are expressed as energy stored by the inductor and capacitor, respectively. As such, the time-averaged electric energy density $u_{e,b}$ stored by the bound charges can be obtained as

$$\langle u_{e,b} \rangle = \frac{1}{4}\left(\rho_L + \frac{1}{\omega^2 \rho_C}\right)|\tilde{\sigma}_b|^2 |\tilde{\mathbf{E}}|^2 \tag{8}$$

where $\tilde{\sigma}_b = j\omega\varepsilon_0\tilde{\varepsilon}_b$ is the complex conductivity for the bound charges. The first term on the right side of Eq. (8) is the energy stored by the inductor (i.e., the kinetic energy of the charges), and the second term is the energy stored by the capacitor (i.e., the potential energy of the charges). Noticing the relation $\rho_L + \frac{1}{\omega^2 \rho_C} = \mathrm{Im}\left[\frac{\partial}{\partial\omega}\left(\frac{1}{\tilde{\sigma}_b}\right)\right]$ or $\left|\frac{\partial}{\partial\omega}\left(\frac{1}{\tilde{\sigma}_b}\right)\right|$, the energy density can be written in the following form,

$$\langle u_{e,b} \rangle = \frac{1}{4}\varepsilon_0 \mathrm{Re}\left[\frac{\tilde{\varepsilon}_b^*}{\tilde{\varepsilon}_b}\frac{\partial(\omega\tilde{\varepsilon}_b)}{\partial\omega}\right]|\tilde{\mathbf{E}}|^2 \quad \text{or} \quad \frac{1}{4}\varepsilon_0\left|\frac{\partial(\omega\tilde{\varepsilon}_b)}{\partial\omega}\right||\tilde{\mathbf{E}}|^2 \tag{9}$$

The second expression in Eq. (9) was first proposed by Vorobyev [11] in which a different approach was used to derive the formula. It is noted there are also other forms of energy density formulae proposed for Lorentz model, such as the formulae presented in by Loudon [5], Ruppin [9], and Shin et al. [10]. Though Eq. (9) is derived for the Lorentz oscillator model based on bound charges, it can also be applied to the Drude model and non-dispersive media as they both can be considered special forms of the Lorentz single oscillator model. The Drude model is a special case with $\omega_0 = 0$, and the non-dispersive media can be considered a special case with $\rho_L = 0$ and $\rho_R = 0$ in the circuit analogy of Eq. (6).

For real materials, the dielectric function model often contains multiple Lorentz oscillators such

that

$$\tilde{\varepsilon} = \sum_i \tilde{\varepsilon}_i = \sum_i \frac{\omega_{p,i}^2}{\omega_{0,i}^2 - \omega^2 + j\omega\Gamma_i} \tag{10}$$

where $\tilde{\varepsilon}_i$ denotes the individual permittivity of the $i^{\text{th}}$ oscillator, which can also be a dielectric term or a Drude term. Based on the additive property of energy, the energy storage in a system of multiple oscillators can be written as

$$\langle u_e \rangle = \frac{1}{4}\varepsilon_0 \sum_i \left|\frac{\partial(\omega\tilde{\varepsilon}_i)}{\partial\omega}\right| |\tilde{\mathbf{E}}|^2 \tag{11}$$

Equation (11) relies only on the value and the first-order derivative of permittivity of individual oscillators at a specific frequency. If the permittivity and its first-order derivative can be described by a few oscillators in a spectral range with good accuracy, then the energy density can be calculated based only on these oscillators since the contributions from other oscillators are negligible. For real materials, it is often difficult or impossible to determine all the oscillators in the full spectral range. The use of a few dominating oscillators provides a practical means for the calculation of the energy density in real materials.

For lossless media with arbitrary number of oscillators (or in the spectral range the loss is negligible), namely, $\varepsilon_i''(\omega) \approx 0$, Eq. (9) is reduced to the well-known formula for lossless dispersive media [8],

$$\langle u_e \rangle = \frac{1}{4}\varepsilon_0 \frac{\partial(\omega\varepsilon')}{\partial\omega} |\tilde{\mathbf{E}}|^2 \tag{12}$$

This formula is useful for calculation of the energy density without relying on the Lorentz oscillator parameters for lossless dielectrics. Fused silica (amorphous $SiO_2$) can be such an example materials in the visible to near-infrared spectral range [22].

It is also noted that mathematically

$$\sum_i \left| \frac{\partial(\omega \tilde{\varepsilon}_i)}{\partial \omega} \right| \geq \left| \frac{\partial}{\partial \omega} \left\{ \omega \sum_i \tilde{\varepsilon}_i \right\} \right| = \left| \frac{\partial(\omega \tilde{\varepsilon})}{\partial \omega} \right| \tag{13}$$

Hence, one obtains

$$\langle u_e \rangle = \frac{1}{4} \varepsilon_0 \sum_i \left| \frac{\partial(\omega \tilde{\varepsilon}_i)}{\partial \omega} \right| |\tilde{\mathbf{E}}|^2 \geq \frac{1}{4} \varepsilon_0 \left| \frac{\partial(\omega \tilde{\varepsilon})}{\partial \omega} \right| |\tilde{\mathbf{E}}|^2 \tag{14}$$

This inequality sets a lower bound for the electric energy density of a general lossy dispersive medium that can be expressed in terms of the relative permittivity. It is useful to estimate the lower limit of the energy density of media in case the exact microphysical model is not known. The relations just presented can be extended to magnetic media if a Lorentz model is assumed for permeability.

The time-averaged overall energy storage density $\langle u \rangle$ includes both electric and magnetic contributions. For a nonmagnetic material it becomes

$$\langle u \rangle = \langle u_e \rangle + \langle u_h \rangle = \frac{1}{4} \varepsilon_0 \sum_i \left| \frac{\partial(\omega \tilde{\varepsilon}_i)}{\partial \omega} \right| |\tilde{\mathbf{E}}|^2 + \frac{1}{4} \mu_0 |\tilde{\mathbf{H}}|^2 \tag{15}$$

## 4. Calculation of local energy density distribution based on RCWA

RCWA is a widely used method and has evolved to be a reliable tool for calculating the diffraction efficiencies of layered periodic nanostructures made of dielectric and/or metallic gratings [27-32]. The procedure for calculating the local energy density distribution and the integrated energy storage in gratings is described in the following. Brenner [33] presented a procedure for calculating the local power dissipation based on RCWA. A similar method is used in the present work and also summarized in this section.

To facilitate the analysis, a dimensionless local energy storage density is introduced as follows:

$$\bar{u}(\mathbf{r}) = \frac{u(\mathbf{r})}{u_0} = \frac{\frac{1}{4}\varepsilon_0 \sum_i \left|\frac{\partial[\omega\tilde{\varepsilon}_i(\mathbf{r})]}{\partial \omega}\right| \left|\tilde{\mathbf{E}}(\mathbf{r})\right|^2 + \frac{1}{4}\mu_0 \left|\tilde{\mathbf{H}}(\mathbf{r})\right|^2}{\frac{1}{4}\varepsilon_0 \left|\tilde{\mathbf{E}}_{inc}\right|^2 + \frac{1}{4}\mu_0 \left|\tilde{\mathbf{H}}_{inc}\right|^2} \qquad (16)$$

where $u(\mathbf{r})$ is the local energy density calculated based on Eq. (15), $u_0$ is the energy density in vacuum based on the incident electric and magnetic fields and can be calculated from $u_0 = \frac{1}{4}\mu_0(\varepsilon_I^{-1}+1)\left|\tilde{\mathbf{H}}_{inc}\right|^2 = \frac{1}{4}\varepsilon_0(1+\varepsilon_I)\left|\tilde{\mathbf{E}}_{inc}\right|^2$, where $\varepsilon_I$ is the relative permittivity of the dielectric medium in the incident region. In the RCWA method [27], the amplitude of the incident magnetic field for TM wave incidence is usually set as unit, namely, $\left|\tilde{\mathbf{H}}_{inc}\right|=1$, and the amplitude of the incident electric field for TE wave incidence is set as unit, namely, $\left|\tilde{\mathbf{E}}_{inc}\right|=1$. As such, for TM wave, the dimensionless local energy density can be calculated from

$$\bar{u}_{TM}(\mathbf{r}) = \frac{\varepsilon_I}{\mu_0(1+\varepsilon_I)}\left\{\varepsilon_0 \sum_i \left|\frac{\partial[\omega\tilde{\varepsilon}_i(\mathbf{r})]}{\partial \omega}\right| \left|\tilde{\mathbf{E}}(\mathbf{r})\right|^2 + \mu_0 \left|\tilde{\mathbf{H}}(\mathbf{r})\right|^2\right\} \qquad (17)$$

For TE wave incidence, the dimensionless local energy density can be calculated from

$$\bar{u}_{TE}(\mathbf{r}) = \frac{1}{\varepsilon_0(1+\varepsilon_I)}\left\{\varepsilon_0 \sum_i \left|\frac{\partial[\omega\tilde{\varepsilon}_i(\mathbf{r})]}{\partial \omega}\right| \left|\tilde{\mathbf{E}}(\mathbf{r})\right|^2 + \mu_0 \left|\tilde{\mathbf{H}}(\mathbf{r})\right|^2\right\} \qquad (18)$$

The total energy storage in a grating layer can be calculated by a volume integration as $U = \int u(\mathbf{r})\mathrm{d}V$. For the convenience of analysis, a dimensionless total energy storage is defined as $\bar{U} = U/(u_0 V)$, where $V$ denotes the volume of the grating layer. For one-dimensional periodic gratings as shown in Fig. 1, the dimensionless total energy storage can be evaluated by

$$\bar{U} = \frac{1}{h\Lambda}\int_0^h \int_0^\Lambda \bar{u}(x,z)\mathrm{d}x\mathrm{d}z \qquad (19)$$

where $h$ is the height of the grating layer and $\Lambda$ is the period of the grating. In this definition, $\bar{U}$ is just the spatial averaged value of the dimensionless local energy storage density $\bar{u}$.

A dimensionless local power dissipation density for nonmagnetic materials is defined as

- 9 -

$$\bar{w}(\mathbf{r}) = \frac{w(\mathbf{r})}{w_0} = \frac{\frac{1}{2}\omega\varepsilon_0\varepsilon''(\mathbf{r})|\tilde{\mathbf{E}}(\mathbf{r})|^2}{\frac{1}{2}\omega\varepsilon_0\varepsilon_\mathrm{I}|\tilde{\mathbf{E}}_{inc}|^2\cos\theta} \tag{20}$$

where $w(\mathbf{r})$ is the local electric power dissipation density calculated based on Eq. (3), $w_0 = |S_{inc,z}|k_\mathrm{I} = \frac{1}{2}\omega\varepsilon_0\varepsilon_\mathrm{I}|\tilde{\mathbf{E}}_{inc}|^2\cos\theta$ is a reference quantity defined based on the incident power flux, in which $S_{inc,z}$ denotes the $z$ component of the time-averaged Poynting vector of the incident wave, $k_\mathrm{I} = k_0\sqrt{\varepsilon_\mathrm{I}}$ is the wavevector of the incident wave in the incident region, $k_0 = \omega/c_0$ is the vacuum wavenumber and $\theta$ is the incident angle (shown in Fig. 1). Equation (20) can be expressed for TM and TE wave incidence, respectively, as follows.

$$\bar{w}_{TM}(\mathbf{r}) = \frac{\varepsilon_0}{\mu_0\cos\theta}\varepsilon''(\mathbf{r})|\tilde{\mathbf{E}}(\mathbf{r})|^2 \quad \text{and} \quad \bar{w}_{TE}(\mathbf{r}) = \frac{\varepsilon''(\mathbf{r})}{\varepsilon_\mathrm{I}\cos\theta}|\tilde{\mathbf{E}}(\mathbf{r})|^2 \tag{21}$$

Similarly, the global power dissipation $W$ can be evaluated by a volume integration $W = \frac{1}{2}\omega\varepsilon_0\int\varepsilon''(\mathbf{r})|\tilde{\mathbf{E}}(\mathbf{r})|^2 dV$. The global power dissipation is better to be expressed in terms of absorptance, which can be calculated from the dimensionless local power dissipation density $\bar{w}(\mathbf{r})$ by

$$\alpha = \frac{W}{|S_{inc,z}|A} = \frac{\frac{1}{2}\omega\varepsilon_0\int\varepsilon''(\mathbf{r})|\tilde{\mathbf{E}}(\mathbf{r})|^2 dV}{\frac{1}{2}\varepsilon_0\sqrt{\varepsilon_I}c_0|\tilde{\mathbf{E}}_{inc}|^2 A\cos\theta} = \frac{k_0\sqrt{\varepsilon_\mathrm{I}}}{A}\int\bar{w}(\mathbf{r})dV \tag{22}$$

where $A$ is the incident area on the surface of the grating. For one dimensional periodic grating, the absorptance can be integrated as

$$\alpha = \frac{k_0\sqrt{\varepsilon_\mathrm{I}}}{\Lambda}\int_{z_\mathrm{min}}^{z_\mathrm{max}}\int_0^\Lambda \bar{w}(x,z)dxdz \tag{23}$$

For multilayer gratings, e.g., the two layer grating illustrated in Fig. 1, the local energy storage and local power dissipation can also be integrated layer by layer to obtain the energy storage and power dissipation of each layer. This is an effective way to recognize the contribution of absorption and storage of energy from each grating layer and it may help analyze the grating-enhanced light trapping in photovoltaic applications [17, 18, 34].

**5. Energy storage and dissipation analysis for the example nanogratings**

In this section, the RCWA method is applied to calculate the local energy density and power dissipation density distribution in two example nanogratings using the formulae described above. The first example is a simple metallic grating and the second is a grating-enhanced thin-film solar cell. The effects of geometry on the characteristics of local energy storage and power dissipation distributions as well as the radiative properties of the nanogratings are analyzed.

5.1 A simple metallic grating or slit array

Consider a simple lamellar grating (or slit array) as shown in Fig. 2(a). The grating is made of silver with a period $\Lambda$ = 500 nm, height $h$ = 400 nm, and slit width $b$ = 50 nm. This grating supports magnetic polaritons (MPs) as demonstrated by Wang and Zhang [35]. The first-order or fundamental mode of MP, MP1, is excited at $\lambda$ = 1.28 μm (wavenumber $\nu$ = 7812 cm$^{-1}$) as shown in Fig. 2(b). The excitation condition of MP1 can be well predicted by an LC circuit model [35]. The dielectric function of silver at room temperature is given by the Drude model $\tilde{\varepsilon}_{Ag}(\omega) = \varepsilon_\infty - \omega_p^2/(\omega^2 - j\Gamma\omega)$ with the following parameters [36]: a high-frequency constant $\varepsilon_\infty = 3.4$, a plasmon frequency $\omega_p = 1.39 \times 10^{16}$ rad/s, and a scattering rate $\Gamma = 2.7 \times 10^{13}$ rad/s. The first and second term in $\tilde{\varepsilon}_{Ag}$ can be considered as two branches of degenerated Lorentz oscillators as mentioned in Section 3.2. MPs in this grating can greatly enhance the absorption and transmission as seen from Fig. 2(b). The transmission and absorption enhancement can be elucidated by investigating the local energy storage and power dissipation in the grating. The energy storage and power dissipation in this grating at the MP resonance condition and another wavelength away from the resonance are compared.

Figures 3 and 4 show the local distributions of dimensionless energy density and dimensionless power dissipation density at two wavelengths: $\lambda = 1.28$ μm ($\nu = 7,812$ cm$^{-1}$) at the MP1 resonance and $\lambda = 0.8$ μm ($\nu = 12,500$ cm$^{-1}$) at a non-resonance wavelength. Only the TM wave at normal incidence is considered since MPs can only be excited by TM wave for this grating. Note that the overall energy density is the summation of electric energy density and magnetic energy density. Besides the overall energy density, the electric and magnetic energy density distributions are also shown. In the plot, the origin of axis is set at the center of the slit on the top surface of the grating. From Fig. 3(a), it is seen that the overall energy density in the slit has a very high value which is generally dozens of times greater than the vacuum energy density at MP1 resonance, as compared to the non-resonance condition shown in Fig. 3(d), though there are still energy density enhancement in the slit, the value of dimensionless energy density is generally less than 5. In general, the energy stored in the slit at MP1 resonance is about an order of magnitude greater than that at non-resonance condition. Further examining at the Figs. 3(b) and 3(c) reveals that the electric energy storage is concentrated mainly at the top and bottom part of the slit, while the magnetic energy storage is concentrated mainly in the middle part of the slit at MP1 resonance. Both electric and magnetic energy densities in the slit are about an order of magnitude greater at MP1 resonance than those at the non-resonance condition. At non-resonance condition, as shown in Figs. 3(e) and 3(f), standing wave pattern of electric energy density and magnetic energy density can be observed in the slit (note that the wavelength of $\tilde{\mathbf{E}}^2$ related to energy density is $\lambda/2$, thus the distance between adjacent nodes is $\lambda/4$). The electric energy density equals magnetic energy density for plane waves in vacuum far away from the nanogratings. Hence the nanogratings not only induce inhomogeneous distribution of overall energy density at different locations, but also alter the ratio of electric energy and magnetic energy at the same location. The local power dissipation is concentrated mainly in a very thin layer of silver wall inside the slit at MP1 resonance as seen from Fig. 4(a). Though the electric energy is concentrated at the top and bottom part of the slit, the local power dissipation occurs mainly at the middle of the slit. When resonance is not excited, as shown in Fig. 4(b), local power dissipation appears both in a thin layer of silver inside the slit and in a thin layer at the upper surface of the silver bar. The distribution also follows the pattern of standing waves. In both cases, the locations in the silver bar where the local absorption is strong also exhibit a strong energy density.

The global radiative properties and the integrated energy storage in the grating layer for the MP1 resonance and non-resonance conditions are presented in Table 1. As can be seen, great enhancement (about an order of magnitude) of absorption and transmission at the resonance is accompanied by the greatly enhanced energy storage in the grating layer. It has also been observed that the enhancement of absorption is usually accompanied by local field enhancements in other subwavelength structures [37-39], indicating the dual appearance of enhancement of absorption (and/or transmission) with energy storage. This phenomenon may be interpreted based on the circuit analogy. The great enhancement often happens at resonance condition, such as surface plasmon resonance or the magnetic resonance. Magnetic resonance can be modeled by an equivalent series RLC circuit [35, 39]. Hence Lorentz model or the equivalent RLC circuit analogy can provide a basic picture of the resonance behavior. The energy is stored by inductors and capacitors in a circuit. It can be seen from the energy storage formula Eq. (8), at the resonance condition, namely $\omega^2 = 1/(\rho_L \rho_C) = \omega_0^2$, there will be $\rho_L = 1/(\omega^2 \rho_C)$, namely, the time-averaged energy storage of the capacitor equals that of the inductor. Furthermore, it is easy to show from Eq. (6) that the impedance of the equivalent circuit reach a minimum at resonance, indicating a maximum current is reached in the circuit. As such, the power dissipation will reach a maximum based on Joule's law. Similarly, the energy stored by the inductor will also reach a maximum. Because the time-averaged energy storage of the capacitor equals that of the inductor at the resonance, the total energy storage will reach a maximum. As such, the condition of maximum power dissipation is also the condition of maximum energy storage. This gives an interpretation of the observation of dual enhancement of power dissipation and energy density in nanogratings. Furthermore, since the energy stored in the capacitor reaches a maximum at resonance condition, the voltage between the two ends of the capacitor will reach a maximum. Noticing the relation that voltage is proportional to the local electric field, this indicates the appearance of a strong electric field in the capacitor at magnetic resonance, such as the strong electric field enhancement observed at the top and bottom of the slit in the example grating shown in Fig. 3(b).

5.2 Grating-enhanced solar cell structure

    The second example considered here is a grating-enhanced thin-film solar cell structure proposed by Wang et al. [34], as shown in Fig. 5(a). The basic thin-film structure without grating is shown in Fig. 5(b) for comparison. For the grating enhanced solar cell, the top layer is a transparent conducting layer made of indium tin oxide (ITO), the second layer is an active layer made of crystalline silicon, the layer under the active layer is a binary silver/ITO grating to induce surface plasmons, and the

bottom layer is the silver substrate that serves as an electrode. The geometric parameters and properties used here are somewhat different from the paper by Wang et al. [33]. The parameters of the structure used in the present study are given as follows: the thicknesses of the ITO layer, silicon layer, and grating layer are $t_1$ = 80 nm, $t_2$ = 100 nm, and $t_3$ = 50 nm, respectively, the grating period $\Lambda$ = 300 nm, and the slit width $w$ = 250 nm. The bottom silver layer is assumed to be thick enough for it to be treated as a semi-infinite medium that is opaque. The dielectric function of crystalline silicon is modeled by four Lorentz oscillators as given in Ref. [40]. The effect of doping on the dielectric function of Si is not considered. A Drude model is used to model the sputtered ITO film [41]. The Drude parameters for silver are the same as in the previous example. While the absorption can also be enhanced by the grating for TE waves, the present work focuses on the surface plasmon polariton (SPP) enhancement only and only TM waves are considered.

The global absorptance ($\alpha$) spectra of the thin-film structure and the grating-enhanced solar cell are shown in Fig. 5(c) for TM waves at normal incidence, in the spectral range from 0.3 to 1.0 μm. Significant global absorption enhancement can be observed for the solar cell structure with grating at $\lambda$ > 0.65 μm. Several peaks can be seen due to SPPs between Si and Ag. The actual resonance conditions or dispersion relations are very complicated due to ITO and the nanostructure effect [33]. The strongest enhancement appears at $\lambda$ = 0.91 μm, where $\alpha$ = 0.098 for the thin-film structure and $\alpha$ = 0.97 for the solar cell with grating, namely, about 10 times enhancement. Figures 6 and 7 show the local distributions of dimensionless energy density and power dissipation, respectively, at $\lambda$ = 0.91 μm for both solar cells. In the plot, the origin of z-axis is set on the top surface of the ITO layer, and the origin of x-axis is set at the center of the slit filled with ITO of the Ag/ITO grating layer. As shown from Fig. 6(a) and 6(d), the overall energy density in the silicon layer is significantly inhomogeneous and concentrated at a relatively small region over the silver bar for the grating-enhanced solar cell as compared to that of the thin-film structure, which is uniform in the silicon layer due to the multilayer structure. Similar scenarios happen for the electric energy and magnetic energy densities as shown in Figs. 6(b) and 6(c) for the grating-enhanced solar cell. This is attributed to the surface plasmon resonance. For the thin-film structure, the electric energy density and the magnetic energy density distributions show standing wave patterns as shown in Figs. 6(e) and 6(f), and the value of energy density is about an order of magnitude smaller than that of the grating-enhanced solar cell. As seen from Figs. 6(a) and 7(a), the local power absorption is concentrated at the location where the electric energy density is strong for the grating-enhanced solar cell, namely, their enhancement are accompanied by each other. Compared to the energy density and power distribution in the thin-film structure shown in Fig. 6(d) and 7(b), both the local energy density and the absorption in the silicon layer are significantly enhanced (about an order of magnitude) in the solar cell with added Ag grating layer, suggesting the global absorption enhancement can be understood as the great enhancement in the local energy density in the silicon layer induced by the Ag grating.

For solar cells, only absorption in the actively layer contributes to photocurrent generation. In order to know the exact amount of electromagnetic energy absorbed in the silicon layer, a layer integration of the local power dissipation is conducted. Table 2 gives the integrated contribution of absorption of each layer (layer absorptance) and the integrated energy storage of each layer both for the thin-film structure and the solar cell with grating. The energy stored in the silicon layer for the grating-enhanced solar cell is enhanced to about 10 times of the thin-film structure. Meanwhile, the absorptance in the silicon layer is also enhanced 10 times. This is similar to the observation in first example that the enhancement of power dissipation is accompanied by the energy storage. Layer absorptance analysis reveals that silicon layer contributes about 71% of global absorption, while the Ag grating layer contributes about 23% of the global absorption in the grating-enhanced solar cell. This indicates that there is still room to improve the grating structure to reduce the useless absorption of the plasmonic grating layer, even though the presented Ag grating has significantly enhanced the absorption of silicon layer. It is thus demonstrated that the method presented here can be very useful for analyzing nanostructure-based solar cells.

## 6. Conclusions

After reexamination of the formulae of the time-averaged energy density, it is noted that there exists a lower bound for the electric energy density of general lossy dispersive media and this lower limit can be related to the overall dielectric function regardless of the number of oscillators and the individual oscillator parameters. The local distributions of energy density and power dissipation density in the nanogratings are investigated by the RCWA method. The information of local energy storage and power dissipation helps to understand the geometric effect on global radiative properties of nanogratings. The enhancement of absorption is accompanied by the enhancement of energy storage both for material at resonance of its dielectric function described by a classical Lorentz oscillator model and for nanostructures at the resonance induced by the geometric structure. At the resonance, the local energy density in nanogratings can be several orders of magnitude higher than the energy density of the incident wave. The appearance of strong local electric field in nanogratings at the geometry-induced resonance is related to the maximum electric energy storage. This study may facilitate the design and optimization of light absorption for photodetectors, solar cells and optical heating devices.

**Acknowledgements**

Work by JMZ was supported by the China Scholarship Council (CSC) and National Natural Science Foundation of China (No. 51121004). Work by ZMZ was supported by the US Department of Energy (Office of Science, Basic Energy Sciences) under Award (DE-FG02-06ER46343).

Table 1. Radiative properties and integrated energy storage in the simple grating at the MP1 resonance ($\lambda$ = 1.28 μm) and a non-resonance wavelength ($\lambda$ = 0.8 μm).

| $\lambda$ (μm) | Absorptance $\alpha$ | Transmittance $T$ | Integrated energy storage in grating $\bar{U}$ |
|---|---|---|---|
| 1.28 | 0.06 | 0.93 | 5.64 |
| 0.8 | 0.01 | 0.03 | 0.41 |

Table 2. The absorptance and integrated dimensionless energy storage in each layer for the grating-enhanced solar cell and the thin-film structure at $\lambda = 0.91$ μm.

| Layers | Grating-enhanced solar cell | | Thin-film structure | |
|---|---|---|---|---|
| | Absorptance | Energy storage | Absorptance | Energy storage |
| ITO | 0.056 | 1.71 | 0.026 | 1.57 |
| Silicon | 0.69 | 14.54 | 0.069 | 1.62 |
| Ag/ITO grating | 0.22 | 6.40 | - | - |
| Silver substrate | 0.004 | - | 0.004 | - |
| Total | 0.97 | 22.65 | 0.099 | 3.19 |

# FIGURE CAPTIONS

**FIG. 1**. Geometry and definitions of basic parameters to illustrate the calculation of energy density and power dissipation using the RCWA. Here, $\Lambda$ is the period of the grating, $h$ is the height of the grating layer, and $\theta$ is the incident angle. Note that only TM wave incidence is considered, and **E**, **H**, and **S** represent the electric field, magnetic field, and the Poynting vector.

**FIG. 2.** (a) Geometry of a simple grating (slit array); (b) the reflectance, absorptance, and transmittance spectra at normal incidence for TM wave incidence. Note that MP1 and MP2 indicate the magnetic polariton resonances.

**FIG. 3**. Distributions of the dimensionless energy densities in the simple grating example: (a) the overall energy density, (b) electric energy density, and (c) magnetic energy density at the resonance wavelength $\lambda = 1.28$ μm ($\nu = 7,812$ cm$^{-1}$); (d) the overall energy density, (e) electric energy density, and (f) magnetic energy density at a non-resonance wavelength $\lambda = 0.8$ μm ($\nu = 12,500$ cm$^{-1}$).

**FIG. 4**. Distributions of the dimensionless power dissipation density in the simple grating example: (a) at the resonance wavelength $\lambda = 1.28$ μm ($\nu = 7,812$ cm$^{-1}$); (b) at a non-resonance wavelength $\lambda = 0.8$ μm ($\nu = 12,500$ cm$^{-1}$).

**FIG. 5.** Illustration of the solar cell structures and their absorptance spectra: (a) grating-enhanced solar cell; (b) basic thin-film structure; (c) the absorptance spectra for TM waves at normal incidence.

**FIG. 6.** Distributions of the dimensionless energy density in the solar cell example at wavelength $\lambda = 0.91$ μm: (a) the overall energy density, (b) electric energy density, and (c) magnetic energy density in the grating-enhanced solar cell; (d) the overall energy density, (e) electric energy density, and (f) magnetic energy density in the thin-film structure.

**FIG. 7.** Distributions of the dimensionless power dissipation density in the solar cell example at wavelength $\lambda = 0.91$ μm: (a) grating-enhanced solar cell; (b) the thin-film structure.

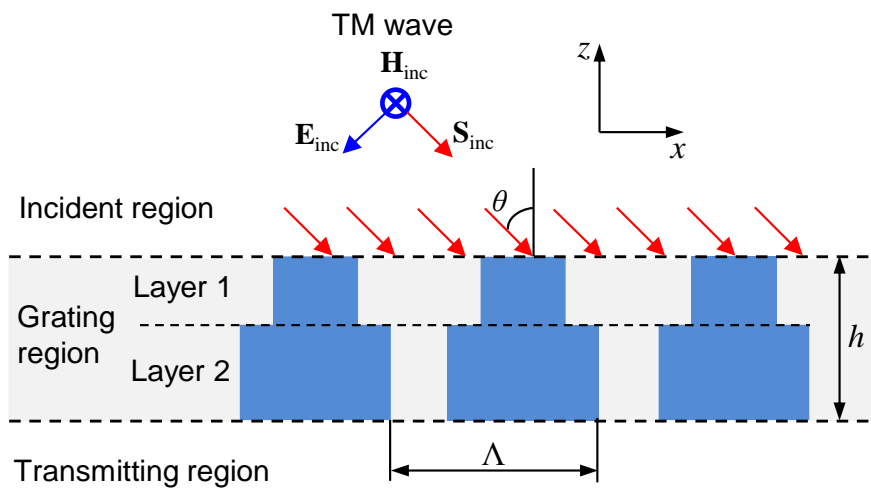

**Zhao and Zhang FIG. 1.**

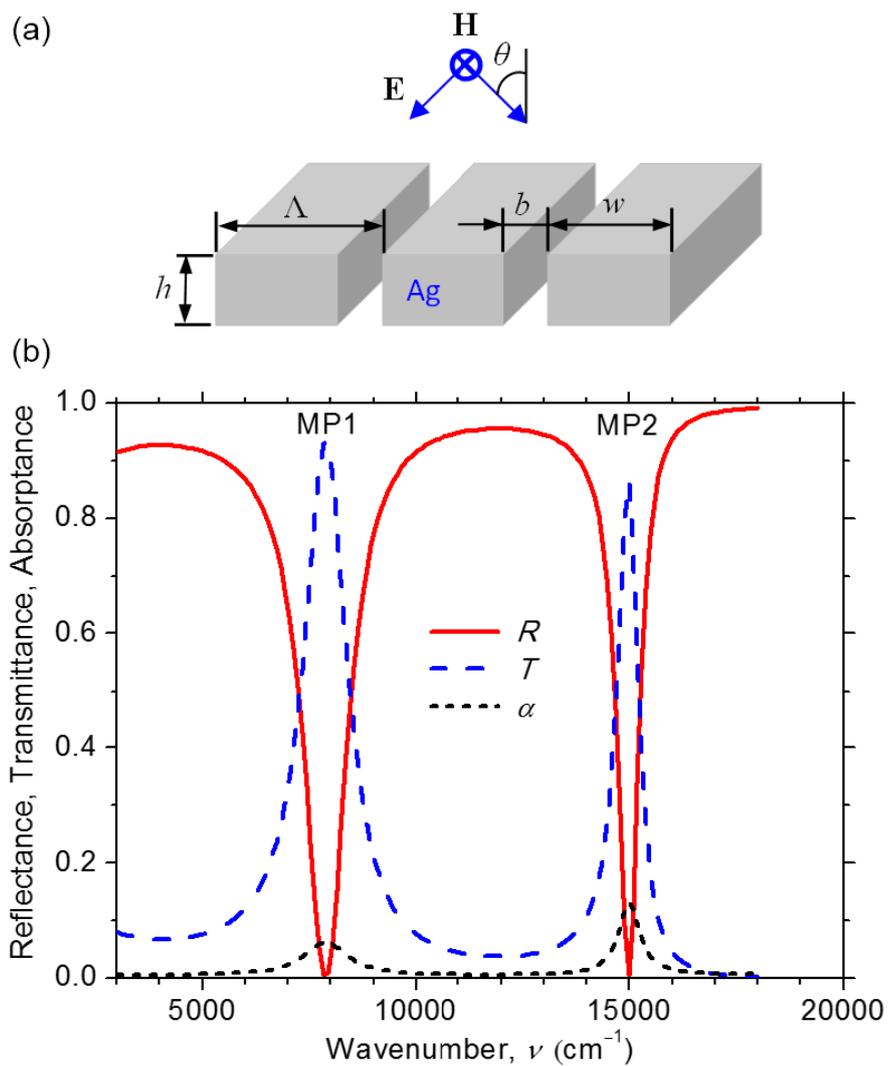

**Zhao and Zhang FIG. 2.**

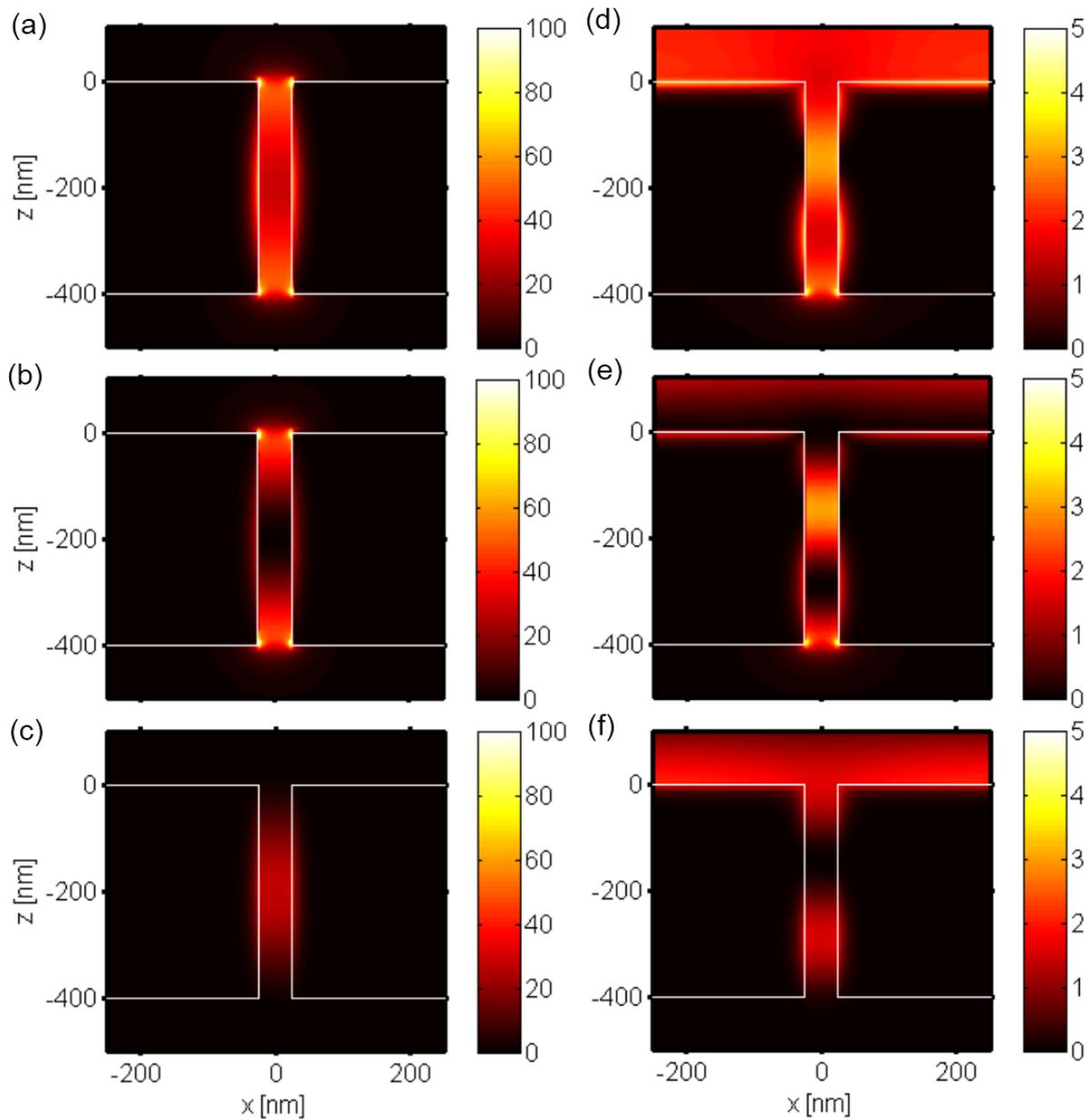

**Zhao and Zhang FIG. 3.**

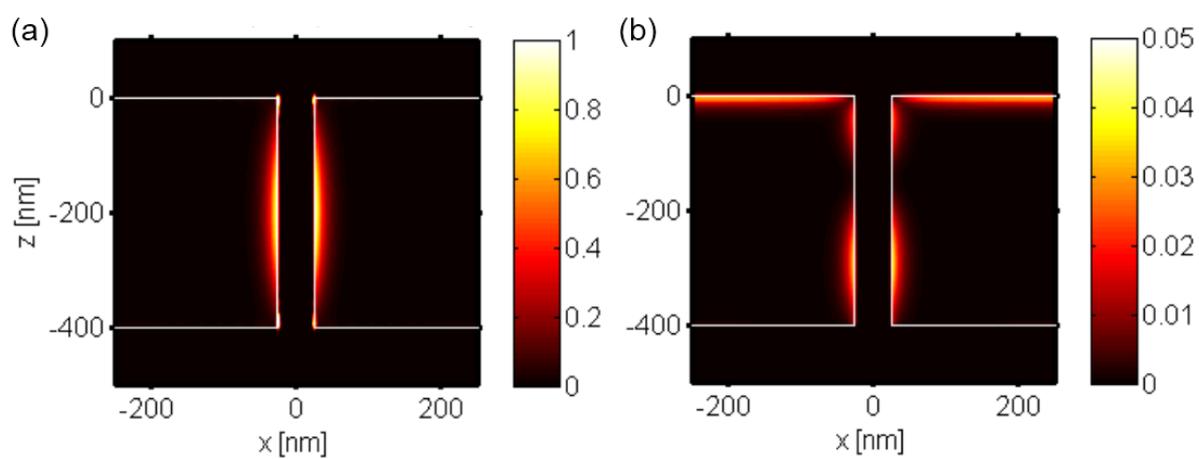

**Zhao and Zhang FIG. 4.**

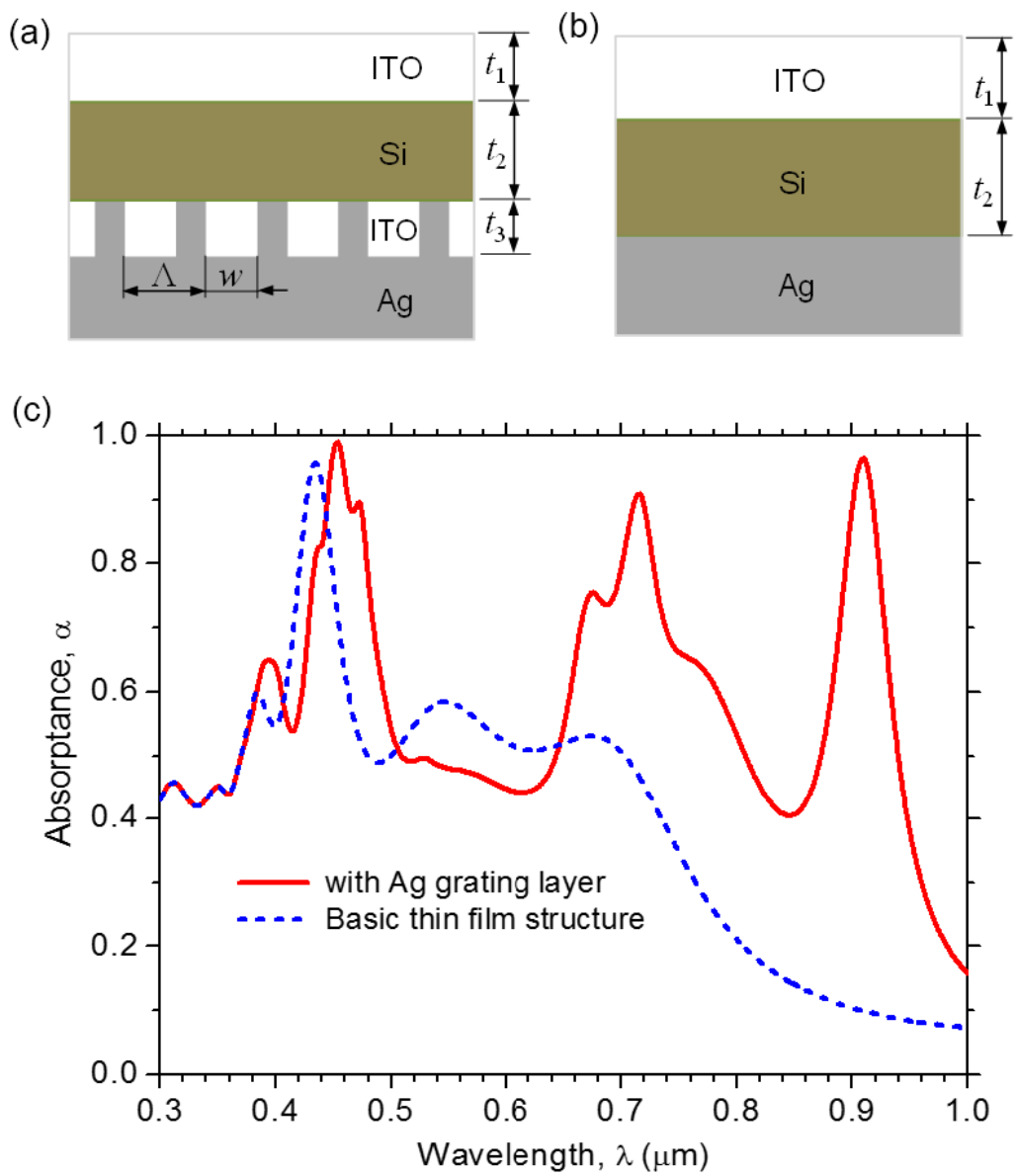

**Zhao and Zhang FIG. 5.**

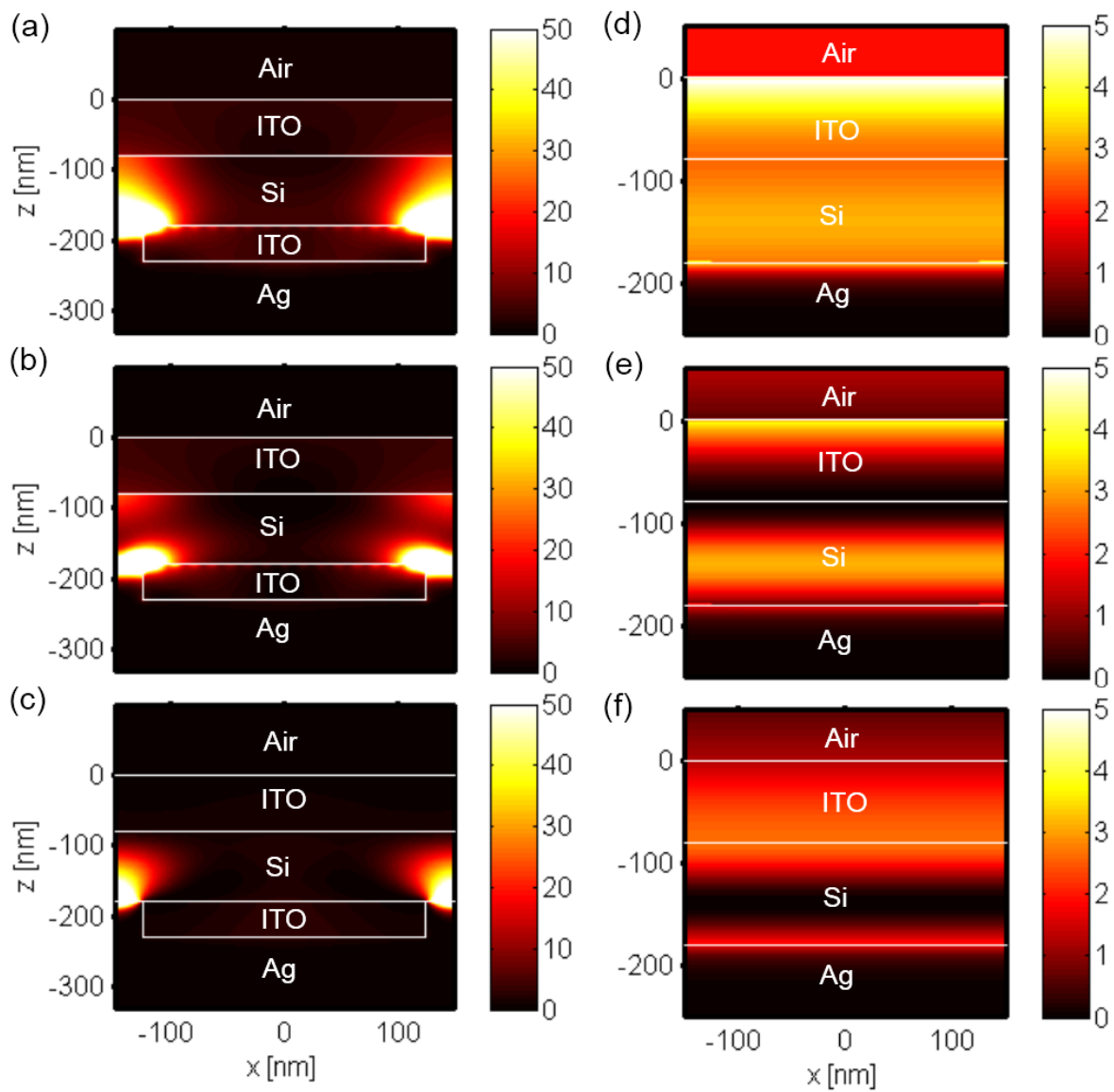

**Zhao and Zhang FIG. 6.**

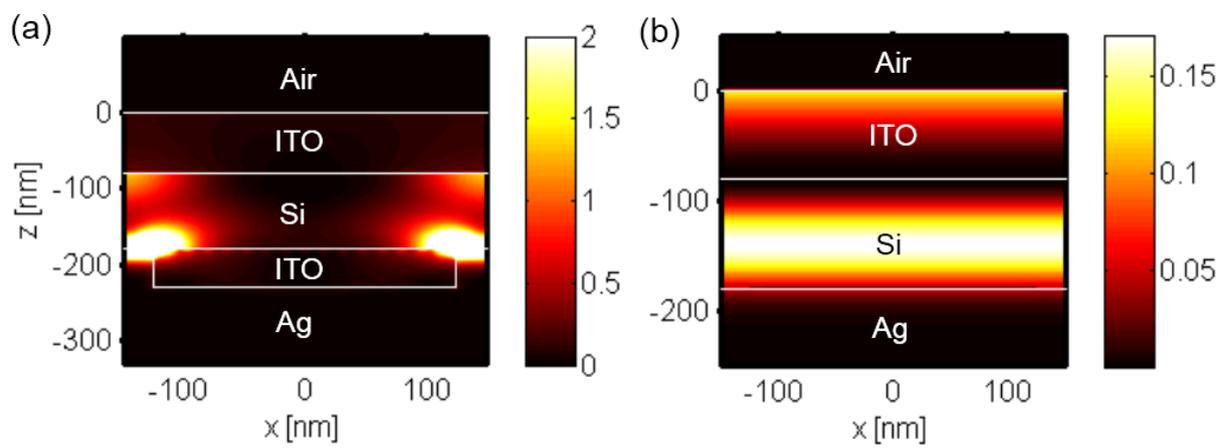

**Zhao and Zhang FIG. 7.**